\documentclass[useAMS,usenatbib]{mn2e}
\usepackage[cp1251]{inputenc}
\usepackage{amsmath}
\usepackage{amssymb}
\usepackage{euscript}
\usepackage[dvips]{graphicx}
\usepackage{epsfig}
\usepackage[usenames,dvipsnames]{color}

\newcommand{\qso}{Q$\,$1232+082}

\title[Partial coverage of the Q~1232+082 BLR by an intervening molecular cloud]
{Partial coverage of the Broad Line Region of Q~1232+082 by an intervening
H$_2$-bearing cloud
\thanks{Based on observations carried out at European Southern Observatory with the Ultraviolet
and Visual Echelle Spectrograph (UVES) mounted on the Very Large Telescope (VLT), unit Kueyen,
               on Cerro Paranal in Chile, under progs. ID 65.P-0038 (P.I. Srianand),
68.A-0106, 69.A-0061, 70.A-0017 (P.I. Petitjean) and 71.B-0136 (P.I. Srianand).}}

\author[Balashev et al.]
{S.A.~Balashev$^{1,2}$\thanks{E-mail: balashev@astro.ioffe.ru},
 P.~Petitjean$^3$, A.V.~Ivanchik$^{1,2}$, C.~Ledoux$^4$, R.~Srianand$^5$,
 \newauthor P.~Noterdaeme$^3$, D.A.~Varshalovich$^{1,2}$\\
\vspace{4pt}\\
$^1$Ioffe Physical-Technical Institute of RAS, {Polyteknicheskaya 26}, 194021 Saint-Petersburg, Russia\\
$^2$St.-Petersburg State Polytechnical University, {Polyteknicheskaya 29}, 195251 Saint-Petersburg, Russia \\
$^3$Universit\'e Pierre et Marie-Curie, Institut d'Astrophysique de Paris, CNRS-UMR7095, 98bis bd Arago, 75014
Paris, France \\
$^4$European Southern Observatory, Alonso de C\'ordova 3107, Casilla
19001, Vitacura, Santiago 19, Chile\\
$^5$IUCAA, Post Bag 4, Ganesh Khind, Pune 411~007, India \\
}
\begin{document}

\date{Accepted 21.07.2011. Received 21.07.2010}

\pagerange{\pageref{firstpage}--\pageref{lastpage}} \pubyear{2009}

\maketitle

\label{firstpage}

\begin{abstract}
We present a detailed analysis of the partial coverage of the \qso\,($z_{\rm em}=2.57$) broad line
region  by an intervening H$_2$-bearing cloud at $z_{\rm abs}=2.3377$.
Using curve of growth analysis and line profile fitting, we demonstrate
that the H$_2$-bearing component of the cloud
covers the QSO intrinsic continuum source completely but only part of
the {Broad Line Region (BLR)}. We find
that only 48$\pm$6~\% of the C~{\sc iv} BLR emission is covered by the
C~{\sc i} absorbing gas.
We observe residual light ($\sim6$\%) as well in the bottom of the
O~{\sc i}\,$\lambda$1302 absorption from the
cloud, redshifted on top of the QSO Lyman-$\alpha$ emission line.
Therefore the extent of the neutral phase of the absorbing cloud is not
large enough to cover all of the background source.
The most likely explanation for this partial coverage is the small size of the intervening cloud, which is comparable to the BLR size.
We estimate the number densities in the cloud: $n_{\rm
H_2}$~$\sim$~110~cm$^{-3}$ for the \mbox{H$_2$-bearing} core
and $n_{\rm H}$~$\sim$~30~cm$^{-3}$ for the neutral envelope. Given the
column densities,
$N$(H$_2$)~=~3.71$\pm$0.97$\times$10$^{19}$~cm$^{-2}$
and $N$(H~{\sc i})~=~7.94$\pm$1.6$\times$10$^{20}$~cm$^{-2}$, we derive
the linear size of the {H$_2$-bearing}
core and the neutral envelope along the line of sight to be $l_{\rm
H_2}$~$\sim$~$0.15^{+0.05}_{-0.05}$~pc and
$l_{\rm HI}$~$\sim$~$8.2^{+6.5}_{-4.1}$~pc, respectively.
We estimate the size of the C~{\sc iv} BLR by two ways (i) extrapolating
size$-$luminosity relations derived from
reverberation observations and (ii) assuming that the H$_2$-bearing
core and the BLR are spherical in shape
and the results are $\sim$0.26 and $\sim$0.18~pc, respectively.
The large size we derive for the extent of the neutral phase of the absorbing cloud together
with a covering factor of $\sim$0.94 of the Lyman-$\alpha$ emission means that the
Lyman-$\alpha$
BLR is probably fully covered but that the Lyman-$\alpha$ emission extends well beyond
the limits of the BLR.
\end{abstract}


\section{Introduction}
\label{introduction}
\noindent
The broad emission lines in the spectra of active galactic nuclei respond
to variations in the luminosity of the central continuum source with a delay due
to light-travel time effects within the emission-line region. It is therefore possible
through the process of 'reverberation mapping' to determine the geometry and kinematics
of the emission-line region by careful monitoring of the continuum variations and
the resulting emission-line response (Blandford \& McKee 1982; Peterson
1993; Netzer \& Peterson 1997).
In particular the size of the broad line region (BLR) can be inferred {from the time delay
 measurement}.
Recent investigations of low-redshift AGNs show a tight relation between this size and
the luminosity of the AGN, $R$~=~$A\times$($L$/10$^{43}$)$^{B}$, where $R$ is the
radius of the BLR, $A$ is a typical distance in light-days and $L$ is the H$\beta$
luminosity in erg/s. The index is found to have a value close to $B$~$\sim$~0.6-0.7 when the
typical distance $A$ is in the range 20-80 light-days (Wu et al. 2004;
Kaspi et al. 2005). Extending this relation to high luminosities yields a typical radius
of the order of 1~pc for the BLR of bright high-$z$ quasars.
The size of the BLR has also been shown to be correlated with the
luminosity in the continuum (Bentz et al. 2009).
%

%
The anti-correlation found between the radius of the region over which an emission line is emitted
and the velocity width of the broad emission line in the same object supports the
idea that the BLR gas is virialized and its velocity field is dominated by
the gravity of the central black-hole (Peterson \& Wandel 1999).
If this is the case, then
the BLR size and the emission line width give an estimate of the mass
of the central object (Peterson \& Wandel 1999; Warner, Hamann \& Dietrich 2003; Wang et al. 2009).
The broad line region is stratified and the
BLR  reverberation mapping size for C~{\sc iv} is about half that for H$\beta$.
This is consistent with the above assumption as
more highly ionized species are expected to be found primarily closer
to the central source of ionization radiation.
%

%
The spatial extent of the BLR is revealed by the partial coverage of some
absorbing clouds, usually associated with the AGN, located in front of the quasar and producing
absorption lines that are saturated but do not go to the zero flux level.
Usually, the continuum source
is covered completely but the emission line region can be covered only partially (e.g. Petitjean, Rauch
\& Carswell 1994, Hamann 1997, {Srianand \& Shankaranarayanan 1999}). In Wampler, Chugai \& Petitjean (1995),
four Fe~{\sc ii} clouds are seen at different velocities with the similar covering factor,
$f\sim 0.5$. In Srianand et al. (2002), line locking and covering factors are shown to be intimately
related and are used to constrain the geometry of the BLR.
Covering factor is one of the characteristics together with variability
and high metallicity that are used to distinguish intrinsic from intervening absorption systems.
Indeed, partial coverage of intervening systems has rarely been reported.
It has been the case in the early Keck spectrum of APM08279+5255 (Ellison et al. 1999; Petitjean et al. 2000a)
which is a lensed quasar whose images are separated by only 0.35~arcsec so that the
Keck spectrum encompasses all the images (Ledoux et al. 1998). It is the case that
the intervening Mg~{\sc ii} systems are not covering all the lensed images (Lewis et al. 2002; Ellison et al. 2004)
and, because of this, typical dimensions of the intervening clouds are derived to be of the order of $\sim$1~kpc.
%

%
Partial coverage of a BLR by an intervening absorber had never been reported before
Ivanchik et al. (2010). These authors note that the C~{\sc i} lines associated
with the $z_{\rm abs}$~=~2.3377 DLA system towards \qso\,\, probably do not cover
the C~{\sc iv} BLR completely so that some flux stays unabsorbed at the bottom of saturated
lines. In the present paper, we analyse in details this unique effect and test
different interpretations. We present the observations in Section~2. Partial coverage
is ascertained in Section~3. Physical conditions of the gas in the DLA are derived
in Section~4 in order to infer {its} extent. Results are discussed in Section~5
before conclusions are drawn in Section~6.

\section{Observations}
\label{observation}
\noindent

The high resolution spectrum of the high redshift quasar
\qso\, ($z_{\rm em}=2.57$ and $m_{\rm V}=18.4$) was obtained
over several observing runs in the course of a survey for molecular
hydrogen in DLA systems with the Ultraviolet and Visible Echelle
Spectrograph (UVES) mounted on the ESO Kueyen VLT telescope on Cerro
Paranal in Chile (Petitjean, Srianand \& Ledoux 2000b; Ledoux et al. 2003,
Noterdaeme et al. 2008). We used Dic1 and central wavelengths 390
and 564~nm in the blue and red arms respectively. The total exposure
time on the source {was} 17.5$\,$h. The CCD pixels were binned
2$\times$2 and the slit width adjusted to 1'' matching the mean
seeing conditions of$\,\sim\,$0.9''. This yields a resolving power
of R$\,\sim\,$50,000 {and S/N ratio varies from 20 in blue arm to 40 in red.} The data were
reduced using the UVES pipeline based on the ESO common pipeline library system. Wavelengths
were rebinned to the vacuum-heliocentric rest frame and individual
scientific exposures were co-added using a sliding window and
weighting the signal by the total errors in each pixel.

The observed QSO spectrum exhibits a DLA at $z_{\rm abs}=2.3377$.
Molecular H$_2$ absorptions associated with the DLA have been detected by
Ge \& Bechtold (1999) and analysed by Srianand et al. (2000) and
Noterdaeme et al. (2008). This is also the system
where the first detection of HD molecules at high redshift was reported
by Varshalovich et al. (2001).

\begin{figure*}
            \includegraphics[width=160mm,clip]{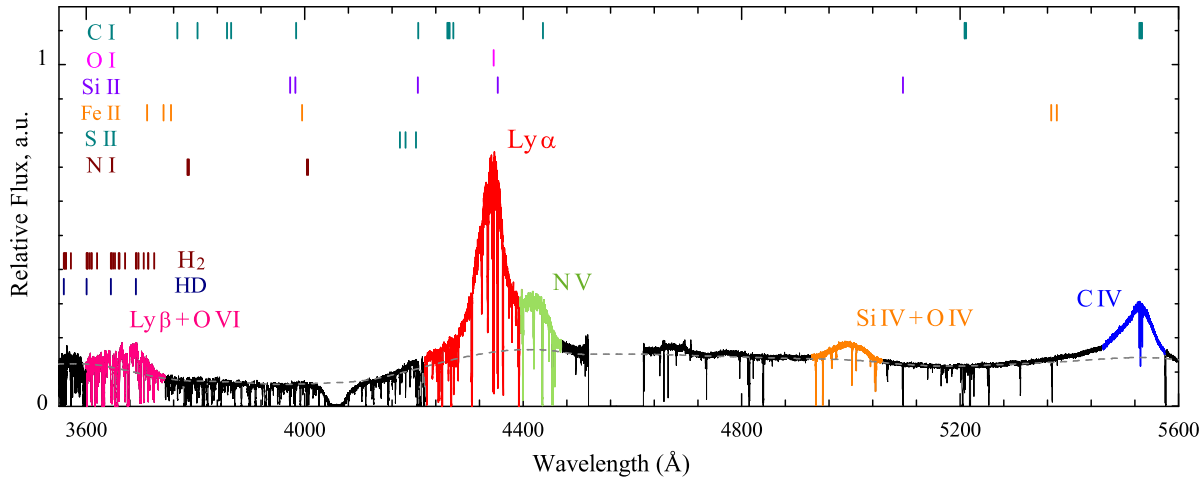}
            \caption{UVES spectrum of Q~1232+082. Emission lines are named and highlighted by different colors, e.g. red for Lyman-$\alpha$,
blue for C~{\sc iv} etc. These colors are used in the following figures to indicate where in the spectrum the corresponding
measurements are performed.
The positions of redshifted absorptions from species discussed in the text are indicated by vertical tick marks.
The gray dashed line shows the Q~1232+082 intrinsic continuum derived by fitting regions devoid of emission and
absorption lines and interpolating the fit below emission lines. This is the continuum we
adopted in the analysis. Note that the spectrum is not flux calibrated.
}
            \label{Coll_Sp}
\end{figure*}

The overall spectrum is shown in Fig.~\ref{Coll_Sp} where we mark the position of
the broad emission lines (highlighted by different colors) as well as the redshifted absorption lines of different species that will be discussed in
the following. The emission lines are Lyman-$\beta$, O\,{\sc vi}$\lambda\lambda$1031,1037, Lyman-$\alpha$, N\,{\sc v}$\lambda$1238,1242, Si\,{\sc iv}$\lambda$1393,1402,
and C\,{\sc iv}$\lambda\lambda$1548,1550. Column densities of H$_2$ and H~{\sc i} in the absorbing cloud are
$N$(H$_2$)~=~3.71$\pm$0.97$\times$10$^{19}$~cm$^{-2}$
and $N$(H~{\sc i})~=~7.94$\pm$1.6$\times$10$^{20}$~cm$^{-2}$ (Ivanchik et al. 2010). \\
In the following, errors are given at the 1$\sigma$ level
and  all column densities are in units of cm$^{-2}$.


\section{Partial coverage}
\label{Partial}

\begin{figure}
            \includegraphics[width=0.45\textwidth,clip]{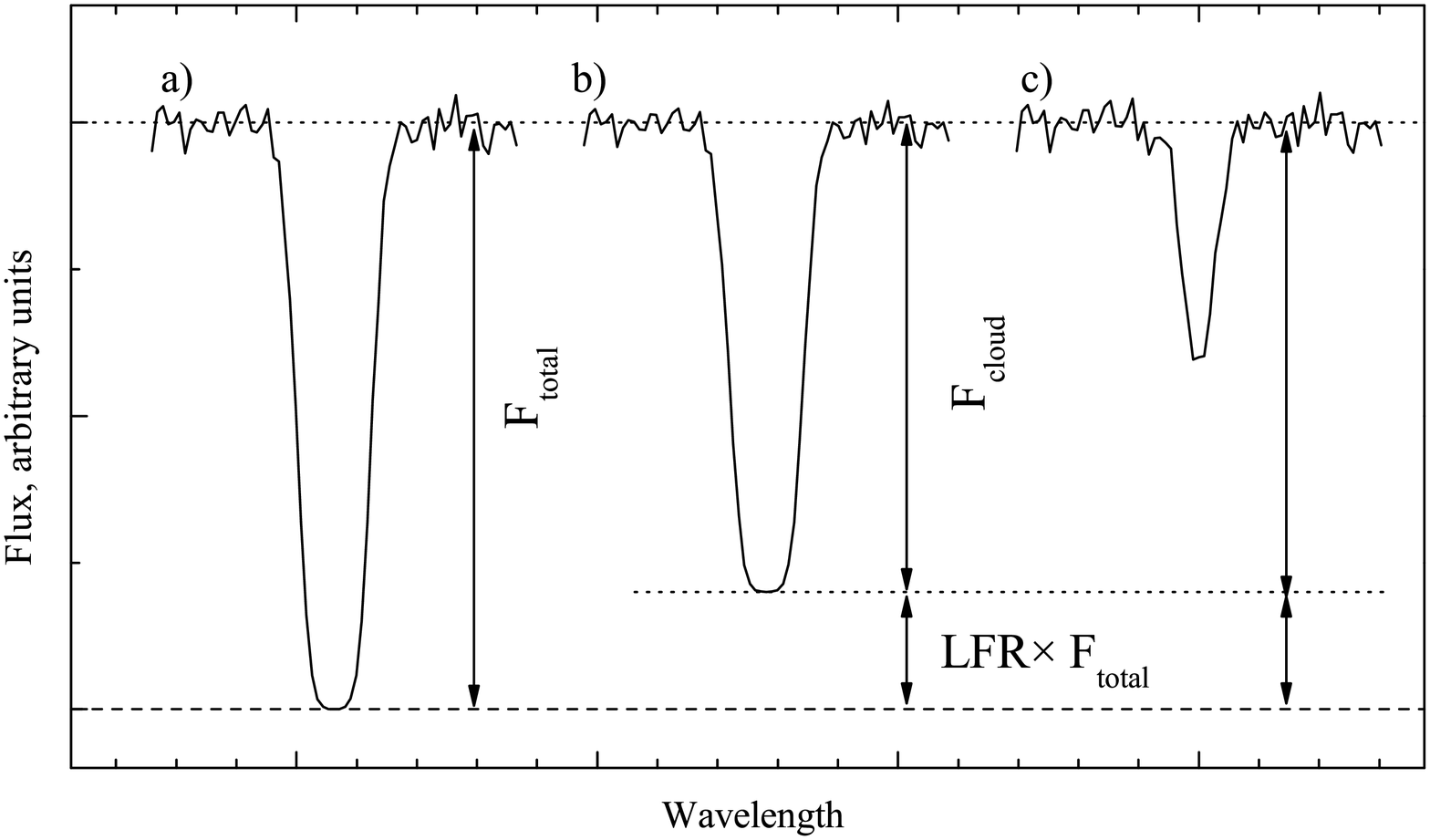}
            \caption{Illustration of the effect of partial coverage on absorption line profiles
            in three cases:
a) a highly saturated line with total coverage, $f=1$; b) a highly saturated
line with partial coverage, $f=0.8$; c) a partially saturated line with partial coverage, $f=0.8$
(see Eq.~2). Line Flux Residual (LFR) is the fraction of
the QSO flux which is not covered by the cloud and can be easily derived in case b).
}
            \label{CovFactors}
\end{figure}

Partial coverage means that only part of the background source is covered by the absorbing cloud. Mainly this
can be the results of (i) the absorbing cloud is smaller than the full projected extent of the background source;
(ii) the absorber, although larger than the background source, is porous, e.g. the filling factor of the cloud is not unity.
This is readily detectable in the spectrum of the background QSO if a saturated line does not go to the zero level indicating that
part of the radiation from the QSO is not shadowed by the cloud. Partial coverage of an emission source by an
absorption system is characterised by covering factor which can be defined as the ratio of
the flux passing through the absorbing cloud, therefore the flux which is affected by absorption, $F_{\rm  cloud}$,
to the total flux, $F_{\rm total}$, which is the continuum flux in the spectrum extrapolated over absorption lines;
\begin{equation}
	f = \frac{F_{\rm cloud}}{F_{\rm total}},
	\label{cf}
\end{equation}
and the measured flux in the spectrum, $F(\lambda)$,  is
\begin{equation}
F(\lambda) = (F_{\rm total}-F_{\rm cloud}) + F_{\rm cloud}(\lambda)e^{-\tau (\lambda)},
\label{flux}
\end{equation}
where $\tau (\lambda)$ is the optical depth of the cloud (see Ganguly et al. 1999). These
definitions are illustrated in Fig.~\ref{CovFactors}.
The determination of the covering factor is trivial in the case of a
highly saturated absorption line (see Fig.~\ref{CovFactors} cases a) and b)). In case of a partially saturated
line (see Fig.~\ref{CovFactors}, c))
several transitions (with possibly different covering factors) must be used.

Absorption lines associated with an absorption system usually span a wide range of wavelengths (see Fig.~\ref{Coll_Sp}).
Therefore we can investigate the dependence of the partial coverage on the position of the lines in the spectrum.
Different species are predominantly found in different regions of the cloud, corresponding to different
physical properties of the gas (molecular, neutral, ionized) and have absorption lines
located on top of different parts of the quasar spectrum attributed to different regions of
the AGN (accretion disk, BLR, NLR).
We therefore can try to infer information on the spatial extent of both the cloud and the AGN.

In the following, we make a detailed analysis of the $z_{\rm abs}=2.33771$ molecular hydrogen absorption system
in the spectrum of \qso\,\, in order to measure the covering factors of different species relative to different
regions of the AGN. We will call Line Flux Residual (LFR) the fraction of the normalized QSO flux which is not covered by the cloud
${\rm LFR} = (F_{total} - F_{cloud})/F_{total} = 1-f $\,(see Fig.~\ref{CovFactors}). In the following LFR is expressed in percents.

\subsection{Zero flux level correction}
\label{correction}

\begin{figure*}
  \includegraphics[width=0.95\textwidth,clip]{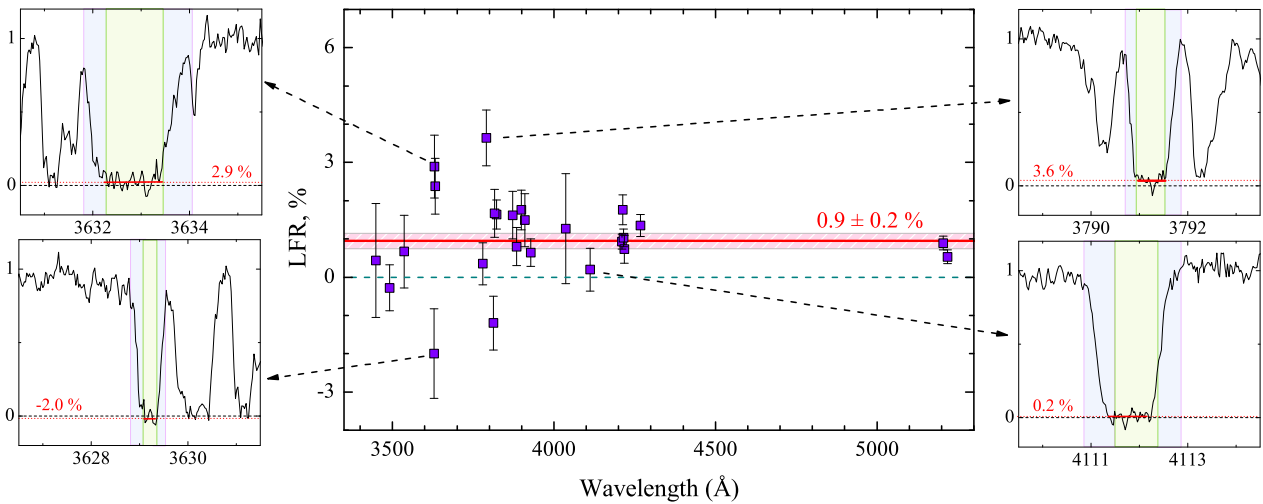}\\
  \caption{Correction of the zero flux level in the spectrum of \qso. {The line flux residual (LFR)} measured in
the core of saturated lines (some of them are shown in individual panels, with green and blue regions indicating
the core and wings of the lines (see text)) that are not associated with the molecular
cloud at $z=2.33771$ is plotted versus wavelength as filled squares. The red horizontal line
shows the mean zero level for these lines. Error bars indicate the scatter in the LFR
measured in individual pixels.}
  \label{Lymancorrection}
\end{figure*}

One possible source of uncertainty in the measurement of partial coverage is
{that the zero flux level of the spectrum is in error
due to approximate background or sky subtraction. }
We have estimated this {uncertainty} by measuring the flux residual at the bottom of saturated
lines located {mainly} in the Lyman-$\alpha$ forest.
Fig.~\ref{Lymancorrection} shows the result of this analysis as the percentage of
the residual flux relative to the flux in the spectrum versus wavelength. We also show
the three lines with the largest error (3.6, 2.9 and 2\%).
Note that a blend of several non-saturated lines could mimic a saturated (broad) line that
does not go to the zero flux level. To avoid as much as possible such lines, we excluded from
the analysis the lines for which the width of the wings (shown as the blue regions in the
individual panels of Fig. \ref{Lymancorrection}) is larger than the full width at half maximum
of the line.

We did not find any systematic dependence of the effective zero intensity level on wavelength.
The average value is found to be $0.9\pm 0.2$\% (see Fig.~2) and we correct the spectrum
for this. Note that the average flux level measured in the core of the DLA absorption line
is found to be $1.8\pm 0.1$\% before correction.


\subsection{{Partial coverage of} H$_2$ absorption lines}
\label{H2pc}

\begin{figure*}
            \includegraphics[width=0.95\textwidth,clip]{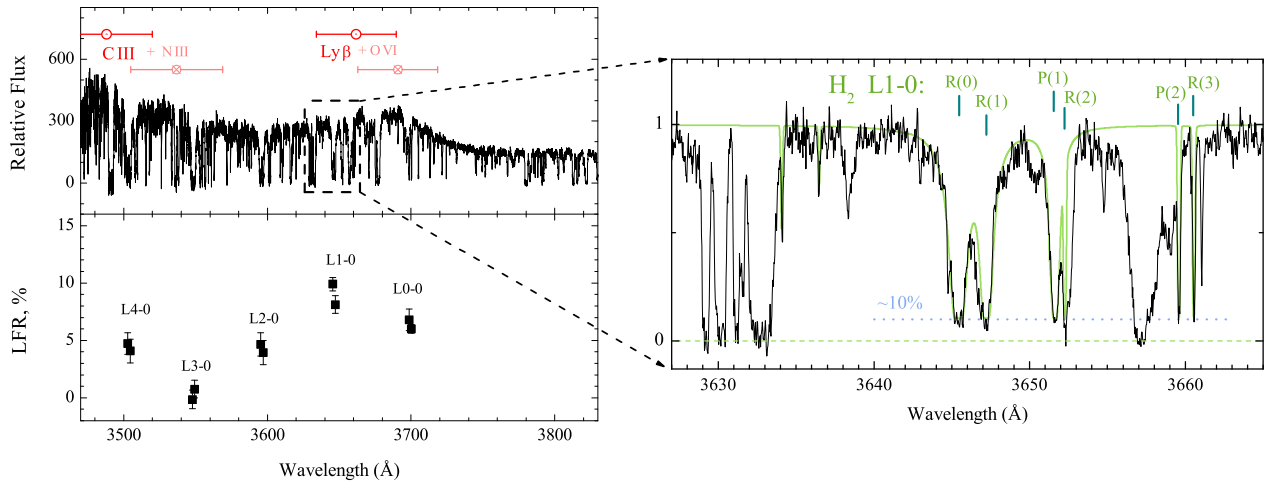}\\
            \caption{{\sl Left top panel}: Part of \qso\,\, spectrum where H$_2$ absorption lines at $z=2.3377$ are located.
Red circles and crosses show the positions of QSO emission lines with error bars indicating line widths taken
from Vanden Berk et al. (2001).
                    {\sl Right panel}: Part of \qso\,\, spectrum showing absorption lines of the L1-0 H$_2$ band
at $z_{\rm abs}$~=~2.33771. The highly saturated and even damped H$_2$ lines do not reach the zero flux level whereas
several saturated intervening H~{\sc i} Lyman-$\alpha$ absorption lines do. Since these lines are redshifted on top of
the Ly$\beta$+O~{\sc vi} emission line from the QSO, this indicates that the {H$_2$-rich} molecular cloud
covers only part of the Broad Line Region.
                    {\sl Left bottom panel}: Line flux residuals (LFR) of H$_2$ absorption lines
Error bars indicate the scatter in the LFR measured in individual pixels.
}
            \label{H2_lines}
\end{figure*}

Because of the presence of prominent lorentzian wings in J~=~0 and 1 H$_2$ transitions,
the H$_2$ column density in these levels can be accurately measured to be $\log N = $  $19.45\pm0.10$
and $19.29\pm0.15$, respectively (Ivanchik et al. 2010).
These column densities imply that the optical depth in the center
of absorption lines is $\gg 1$, i.e. all absorption lines from the J~=~0 and 1 levels are highly saturated.
In addition the widths of J~=~0 and 1 lines are larger than the instrumental broadening, therefore all J~=~0 and 1
absorption lines should reach zero flux level in their center.
Nonetheless, it can be seen in the right panel of Fig.~\ref{H2_lines} that the profiles of these strongly saturated
H$_2$ absorption lines do not go to the zero flux level while the nearby saturated Lyman-$\alpha$ lines do
(with an uncertainty $<2$~\%, see Sect.~\ref{correction}).
The residual flux in the H$_2$ absorption lines reaches $\sim$10~\% of the QSO flux at the corresponding positions.


The large number of the H$_2$ absorption lines and the fact that they are located in the
same portion of the spectrum, provides a good opportunity to investigate more the covering factor
of the H$_2$-bearing cloud.
{Only H$_2$ absorption lines from the J~=~0 and 1 rotational levels are strongly saturated and used to measure
the covering factor.}
We measure the LFR as the average of the flux residuals in the pixels of the
bottom of the saturated lines. The corresponding values are shown as filled squares
in the left-bottom panel of Fig.~\ref{H2_lines}.

There is a hint for the LFR values to be larger on top of emission lines. This supports the idea that the
{H$_2$-bearing} cloud covers the central source of continuum but does not cover the whole BLR.
The position of these lines are roughly indicated as horizontal segments in Figure~\ref{H2_lines}. Positions and widths of the
QSO emission lines are taken from Vanden Berk et al. (2001). Note that Lyman-$\beta$ is blended with O\,{\sc vi} and C\,{\sc iii}
with N\,{\sc iii}. Corresponding emissions are clearly seen in the \qso\,\, spectrum (see upper left panel in Fig.~\ref{H2_lines}).
Note that H$_2$ L3-0 is blended with intervening Lyman-$\alpha$ lines.


\subsection{{Partial coverage of} C\,{\sc i} absorption lines}
\label{CIpc}

Neutral Carbon absorption lines from the three fine structure levels of the ground state
are seen in one single component associated with the {H$_2$} molecular system.
These levels are denoted in the following as C\,{\sc i} (ground state), C\,{\sc i}$^{*}$ (23.6\,K above the ground state),
and C\,{\sc i}$^{**}$ (62.4\,K above the ground state).
The lines from these levels are not highly saturated, span a wide wavelength range in the QSO spectrum $\approx$3800$-$5600~\AA~
(see Fig.~\ref{Coll_Sp}) and a wide range of oscillator strengths. In addition, by chance, some of these lines are located on top
of the QSO C\,{\sc iv} emission line. C~{\sc i} and H$_2$ are believed to be nearly co-spatial in {diffuse} molecular
clouds {(Srianand et al. 2005)}
and therefore we can expect the C\,{\sc i} lines to show partial coverage as well.
To estimate the LFR of these lines, we use two different methods, curve-of-growth and profile fitting.


\subsubsection{Curve of growth}
\label{CurveOfGrowth}
\noindent

Atomic data for C\,{\sc i} transitions
were taken from the {Wiese, Fuhr \& Deters (1996)}. Only lines without any apparent blend with absorptions from
other species were used in the analysis. The spectrum was normalized with a continuum constructed by fitting spline
functions to points devoid of any absorption.

Some of the lines partly overlap with each others because of close
wavelengths. If the overlapping lines belong to different C\,{\sc i} fine-structure levels
(e.g. $\lambda_0 \approx 1277.2$~\AA), then the equivalent width (EW) was measured by fitting
simple absorption components with fixed relative velocity shifts (but different central
optical depth). If the {overlapping} lines belong to the same fine-structure level (as for $\lambda_0 \approx 1329.1$~\AA),
then we fitted a profile where all the parameters were fixed except one for the central
optical depths.

\begin{figure*}
            \includegraphics[width=150mm,clip]{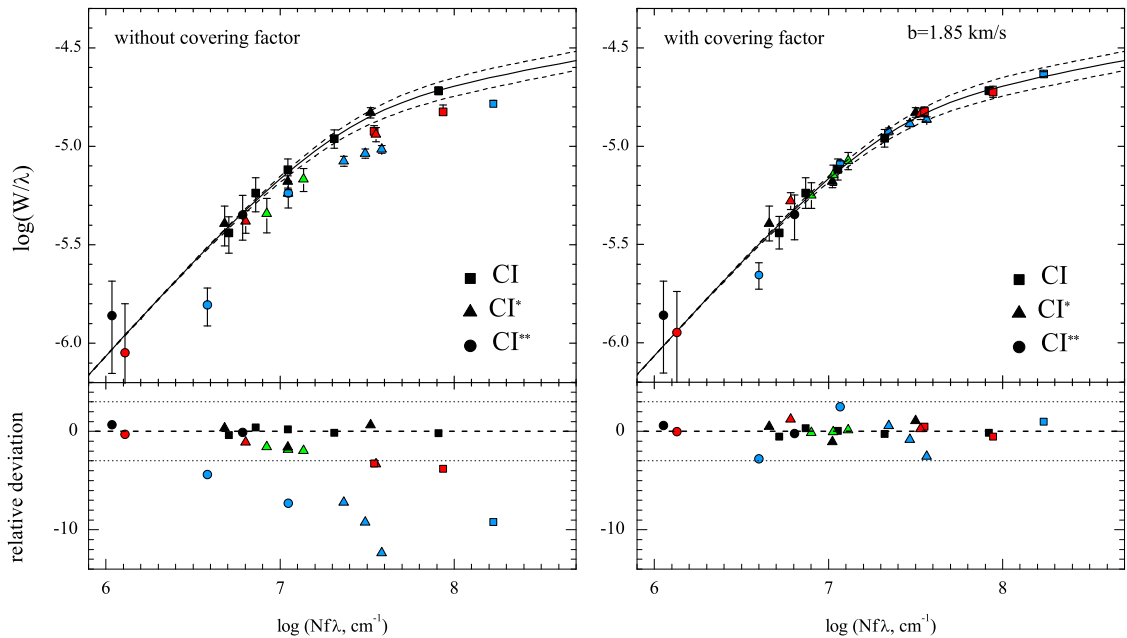}
            \caption{Comparison of the curves-of-growth obtained in two cases:
{\sl Left hand-side panel}: using only absorption lines located on top of the
QSO intrinsic continuum {(black points)}. In that case, absorption lines located on top of emission lines
{(colored points) significantly} deviate from this curve;  {\sl Right hand-side panel}: using all the lines
but correcting for partial coverage (see Table~1). The error bars represent 1$\sigma$ error determination
of the equivalent widths. The dashed lines show the theoretical curve of growth with a Doppler parameter
deviating from 1$\sigma$ interval for estimated b-parameter for the best curve-of-growth.
Bottom panels show relative deviations of measured equivalent widths from the best curve of growth. Relative deviation
of a point is the deviation measured in units of obtained errors for this point. The dotted curves show relative
deviations equal to $\pm3$.
}
            \label{cog_comparison}
\end{figure*}

We construct the observed curve-of-growth for C\,{\sc i} absorption lines detected in the spectrum (see
left panel of Fig.~\ref{cog_comparison}). In this figure, black points are for absorption lines
located on top of the QSO intrinsic continuum whereas other color points are for absorption lines
located on top of an emission line with the color code as indicated on Fig.~1.
The best theoretical fit to the black points yields $b=1.85$~km/s. It can be seen in the left panel
of Fig.~\ref{cog_comparison} that not only this fit is very good but that most of the other
points do not lie on the theoretical curve. These lines have a too small equivalent width.

The large discrepancy between the observed equivalent width and the one expected from the
theoretical curve fitted on the lines located on top of the intrinsic quasar continuum is maximum for
the blue subset of lines that are located on top of the C\,{\sc iv} emission line. It
cannot be explained by continuum misplacement since the signal to noise ratio in this part of the spectrum is
quite large ($\sim$50). Blending with other absorption lines also cannot explain this discrepancy because blends
tend to increase the equivalent widths while we see systematically smaller equivalent widths.
On the other hand partial coverage is a reasonable explanation for this discrepancy because it
means that part of the background radiation is not intercepted by the absorption system yielding a decrease of the measured equivalent width,
$W_{\rm obs} = W(N,b)\cdot f$, where $N$ and $b$ are, respectively, the column density
and the Doppler parameter, and $f$ is a covering factor.


We re-fitted the curve of growth with additional minimization parameters
corresponding to the LFR values. Since the C\,{\sc i} lines are grouped around similar wavelengths,
we applied {\sl the same value of LFR for lines that are members of a group}. Therefore different LFR values are used for
the groups located on top of the C~{\sc iv}, Lyman-$\alpha$ and N~{\sc v} emission lines, respectively (see Fig.~\ref{Coll_Sp}).
The lines located on top of the QSO intrinsic continuum are assumed to fully cover the background source.
Our best fit using the new $\chi^2$ procedure is presented in the right panel of Fig.~\ref{cog_comparison}
and the best fitted parameters are given in Table~\ref{table_res}. The derived LFR values are of the order of 20-30~\%.
The LFR error is found to be larger for the lines located on top of the N~{\sc v} emission line because of larger
errors in the continuum placement in this region. The reduced $\chi^2$-value after correction for partial coverage is $\approx$1.2 instead of
$\approx$3.8 for the case without correction. The $b$ values and C\,{\sc i} column densities for the final fit
and the fit of the absorption lines that are located on top of the continuum are in agreement which indicates that this partial
coverage explanation
is quite satisfactory.


\subsubsection{Voigt-profile Fitting}
\label{ProfileFitting}
\begin{figure*}
      \includegraphics[width=1.0\textwidth,clip]{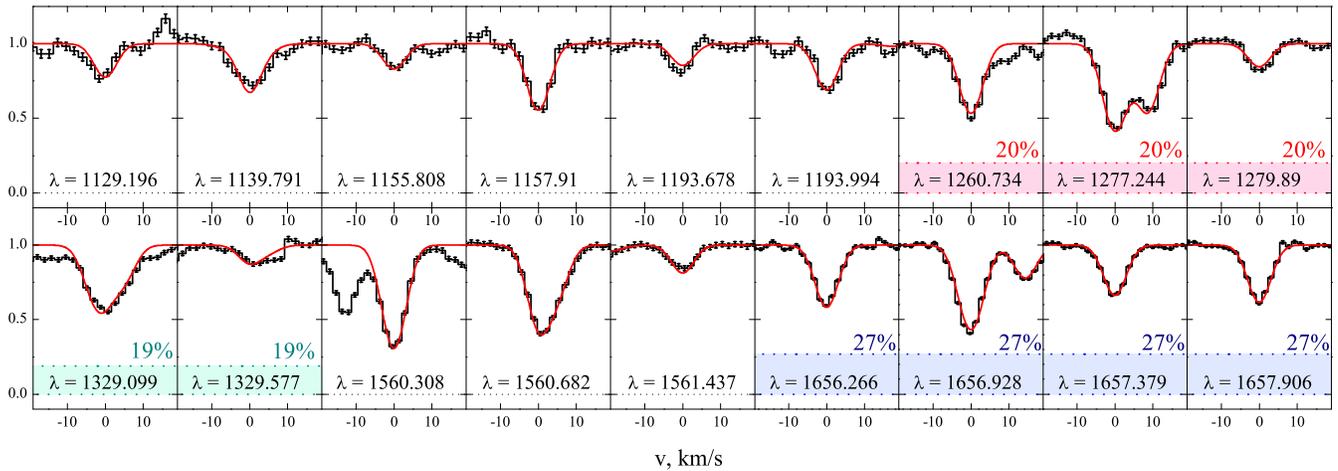}
      \caption{Final Voigt-profile fit of the C~{\sc i} absorption lines.
The overall fit is overplotted as a red solid line.
Partial coverage of the emission lines (indicated in each panel) is derived from the fit and
is found to be $\sim$20~\% for Lyman-$\alpha$, 19~\% for N~{\sc v} and
27~\% for C~{\sc iv} (see Table~1).
Colors correspond to the location of the lines in the spectrum as
shown in Fig.~\ref{Coll_Sp}.
}
            \label{synth_spectrum}
\end{figure*}

\noindent
We performed Voigt-profile fitting of C~{\sc i} absorption lines to confirm the results obtained
from the curve-of-growth analysis. As before, the continuum was locally approximated by spline functions.
 We fitted the absorption lines adding to usual parameters ($b$ and $N$)
a covering factor parameter for each of the three main emission lines of the QSO spectrum
(C~{\sc iv}, Lyman-$\alpha$ and N~{\sc v}) to be constrained during the minimization.
A few line profiles together with the corresponding fitted spectrum are shown in Fig.~\ref{synth_spectrum}.
In each panel we indicate the required LFR, $\sim$20, 19 and 27~\%, respectively, for
Lyman-$\alpha$, N~{\sc v} and C~{\sc iv} (see Table~1).

\begin{figure}
       \includegraphics[width=0.48\textwidth,clip]{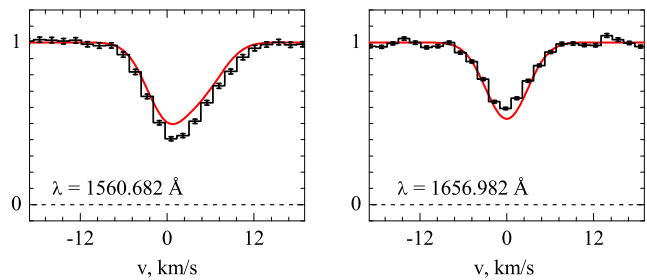}
       \caption{Examples of poorly fitted lines when the fit to all C~{\sc i} transitions
is performed without taking into account partial coverage. Left panel shows the case of a
line which is not affected by partial coverage. In that case, the fit underestimates
the profile because the overall fitted column density is found too small due to the weight of lines affected
by partial coverage.
Right panel shows the case of a transition affected by partial coverage. In that case the fit overestimates
the absorption because the partial coverage decreases the observed optical depth of the transition.
Note that the effect is large in the present case because LFR is found to be as large as 27~\%
for the lines on top of the C~{\sc iv} emission line (see Fig~6).
}
            \label{ss_example}
\end{figure}

The final reduced $\chi^2$ is $\approx$3 when it is $\approx$10
without any correction of partial coverage.
Fig.~\ref{ss_example} demonstrates that partial coverage correction
is indeed needed. Partial coverage makes a line with column density $N$ and therefore
optical depth $\tau_{\rm 0}$ look like a line with $\tau_{\rm obs}$~$<$~$\tau_{\rm 0}$.
When fitting with the same $N_{\rm o}$ a series of absorption lines, some of which are affected by
partial coverage (when on the top of emission lines) and some of which are not affected
by partial coverage (when on the top of the intrinsic continuum), the value of $N$ derived from
the fit will be smaller than $N_{\rm o}$. Therefore, the fit of the lines affected by
partial coverage will be deeper than the absorption line whilst the fit of the lines
that are not affected by partial coverage will not be deep enough.
This is indeed the case for the $\lambda$1560.682 (not affected)
$\lambda$1656.982~\AA~ (affected) features as can be seen on Fig.~\ref{ss_example}.

Note that a covering factor parameter was introduced as well for absorption lines
located on top of the QSO intrinsic continuum and the best fit is obtained with total coverage of these lines.

Results of profile fitting are shown in Table~\ref{table_res}.
Parameters obtained from curve-of-growth analysis and profile fitting are in excellent agreement {with each other}
and both slightly different from results obtained previously without consideration
of partial coverage by { Srianand et al. (2005)}.


\begin{table*}
            \begin{tabular}{|c|c|c|c}
            \hline \hline
            value & ~~~~{Curve of growth}~~~~ & ~~~~Profile fitting~~~~ &  {Srianand et al. 2005}\\
            \hline
            \hline
            $b$ (km/s)                              & ~1.85 $\pm$ 0.09  & ~1.84 $\pm$ 0.15 & ~1.70 $\pm$ 0.10 \\
            ~log~$N$(C\,{\sc i})        & 13.87 $\pm$ 0.05  & 13.87 $\pm$ 0.05 & 13.86 $\pm$ 0.22 \\
            ~~log~$N$(C\,{\sc i}$^*$)     & 13.58 $\pm$ 0.04  & 13.56 $\pm$ 0.04 & 13.43 $\pm$ 0.07 \\
            ~~~log~$N$(C\,{\sc i}$^{**}$) & 12.83 $\pm$ 0.05  & 12.82 $\pm$ 0.07 & 12.63 $\pm$ 0.22 \\
            \hline
            LFR: C\,{\sc iv}               &  ~29.2 $\pm$ 3.2 \% & ~26.9  $\pm$ 4.4 \% &  \\
            LFR: Ly$\alpha$                &  ~20.9 $\pm$ 4.8 \% & ~19.6  $\pm$ 5.3 \% &  \\
            LFR: Ly$\alpha$ + N\,{\sc v}   &  ~19.1 $\pm$ 6.7 \% & ~19.4  $\pm$ 8.8 \% &  \\
            \hline
            \hline
            \end{tabular}
            \caption{Column densities and Doppler parameters obtained from curve-of-growth analysis and
Voigt profile fitting, compared to the results without considering partial coverage ({Srianand et al. 2005}). Column densities are in units of cm$^{-2}$.
}
            \label{table_res}
\end{table*}


\subsection{Si$\,${\sc ii} and O~{\sc i}}
\label{SiandO}

\begin{figure*}
            \includegraphics[width=0.9\textwidth,clip]{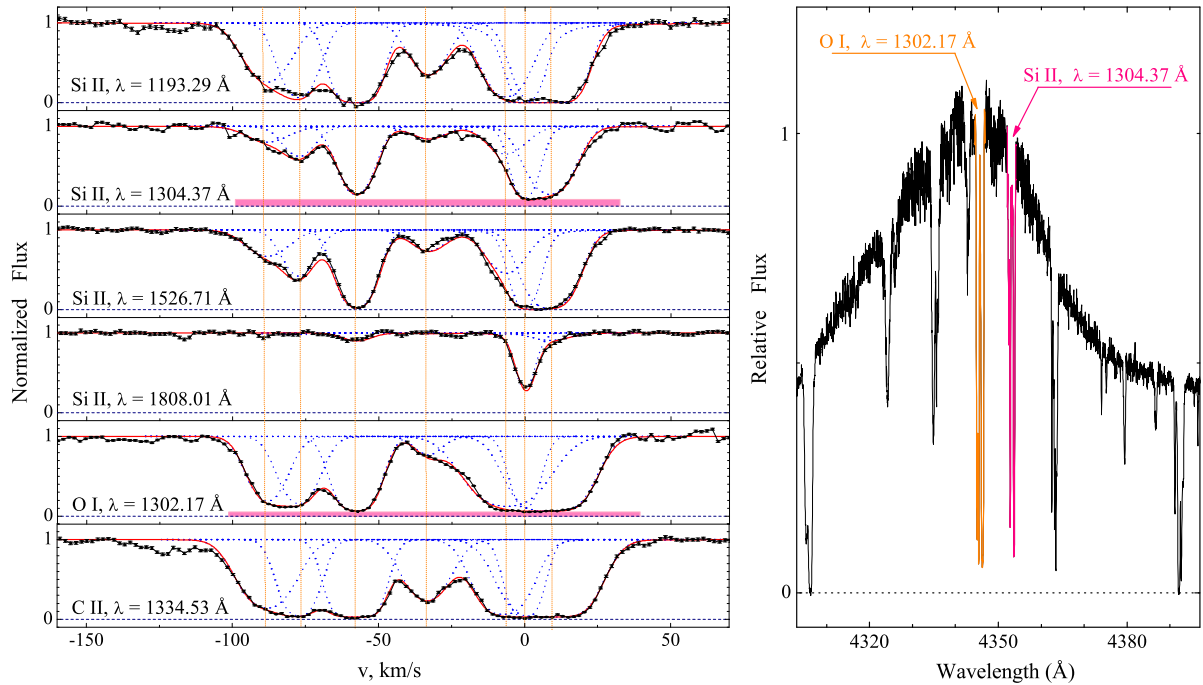}
            \caption{Si\,{\sc ii}, O\,{\sc i} and C\,{\sc ii} absorption features at
$z=2.3377$ in the \qso\,\, spectrum.
                    Results from Voigt profile fitting together with components are overplotted
in the left hand side panel. The velocity origin is taken at the position of the {H$_2$-bearing component}.
                     The right panel shows the location of the
O~{\sc i}\,\,$\lambda$1302 and Si~{\sc ii}$\,\,\lambda$1304 absorption lines on top of
the QSO Lyman-$\alpha$ emission line together with two saturated lines going to the zero level
on both sides.
}
            \label{SiandOfig}
\end{figure*}

In addition to the C~{\sc i} and H$_2$ transitions studied above, we found two additional absorption lines
associated with the $z_{\rm abs}=2.3377$ absorption system and located on top of the QSO Lyman-$\alpha$
emission line that show also partial coverage: \mbox{$\lambda = 1302.2$ \AA} of O\,{\sc i} and \mbox{$\lambda = 1304.4$ \AA}
of Si\,{\sc ii} (see Fig.~\ref{SiandOfig}).
The main component of the O\,{\sc i}\,$\lambda$1302 transition at $v\sim 0$~km/s, has an apparent flat bottom
but with non-zero flux at the center of line. This feature cannot be the result of finite spectral resolution as its width
is about 50 km/s, much larger than our resolution, FWHM~$\sim6$~km/s.
Also the shape is very unlikely to be a combination of many unsaturated lines because all the components, the structure
of which can be guessed from Si~{\sc ii}, would have to have exactly the same optical depth.
Thus, assuming that there is no error in the zero flux level as ascertained by two saturated lines on both side of the Lyman-$\alpha$
emission line, the LFR for this line is $\sim$6~\%.

We detect six Si\,{\sc ii} transitions at $z=2.3377$.
One of them (Si\,{\sc ii} $\lambda$1304) is located on top of the Lyman-$\alpha$ emission line as well and is expected to
show partial {coverage} as O~{\sc i} $\lambda$1302. The others are located on top of the continuum only.
We performed a joint fit of the Si\,{\sc ii} and O\,{\sc i} lines, using seven components (one of them found coincident with
the molecular component at $z=2.33771$) and allowing for partial coverage for O~{\sc i}$\lambda$1302
and Si\,{\sc ii}$\lambda$1304. We have tied up together the velocity positions of the seven components
and let the other parameters vary (Doppler widths and column densities). We find the values of LFR for these lines
are 6.2~\% and 7.5~\% for O~{\sc i} and Si~{\sc ii}$\lambda$1304 respectively.
We have minimized the number of free parameters
using one LFR value for all components of each absorption profile: O~{\sc i}$\lambda$1302 \AA\, and Si\,{\sc ii}$\lambda$1304 \AA.
The O~{\sc i} component at $-80$~km/s could have LFR slightly larger than the main component at $v=0$~km/s. However, the
width of the component is only $\sim$20~km/s and the bottom of the line is not perfectly flat. Therefore, we have no strong argument
to claim that LFR is larger for this component.

We have fitted the Si~{\sc ii} and O~{\sc i} absorption features together with
the corresponding Al\,{\sc ii}$\lambda$1670 and C\,{\sc ii}$\lambda$1334 transitions.
The results of the fit is presented in the left-hand side panel of Fig.~\ref{SiandOfig}.
where LFR for Si~{\sc ii}$\lambda$1304 and O~{\sc i}$\lambda$1302 are illustrated by red stripes.
Note that the possibility is left to have a LFR of $\approx$2~\%  for each of the Al\,{\sc ii}$\lambda$1670 and
C\,{\sc ii}$\lambda$1334 transitions which are located in the far wings of the N\,{\sc v} and C\,{\sc iv} emission
lines, respectively. We find however that this value is too close to the average error in the zero flux level to be
confident it is real.


\subsection{Other {species}}
\label{Otherpc}

\begin{figure}
            \includegraphics[width=0.45\textwidth,clip]{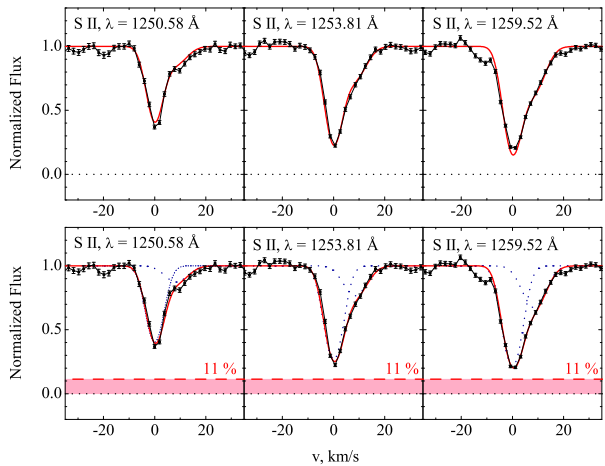}
            \caption{S\,{\sc ii} absorption lines associated with the $z=2.3377$ absorption system.
            The results of two variants of a two component fit are presented. The top panels are for full
coverage and the lower panels for a LFR of 11\%.
}
            \label{S_II}
\end{figure}

Numerous other absorption lines from Fe\,{\sc ii}, S\,{\sc ii} and N\,{\sc i} are present in
the spectrum. N\,{\sc i} transitions are located in a wavelength range devoid of QSO emission lines
while some lines of Fe\,{\sc ii} and S\,{\sc ii} fall partially in the wings of emission lines (see Fig.~\ref{Coll_Sp}).
These absorption lines are not highly saturated and have similar line strengths,
which makes the determination of the covering factor unreliable.
For example, the three S\,{\sc ii} lines are located in
the wing of the QSO Lyman-$\alpha$ emission line.
The profile fitting analysis of the three lines indicates that a LFR of 11$\pm$5\%
is slightly preferred to full coverage (see Fig.~\ref{S_II}).
A standard statistical analysis yields a reduced $\chi^2$ of 2.8 when partial coverage is
allowed instead of 3.7 with full coverage. We think however that the conclusion of
partial coverage for the S~{\sc ii} lines is only tentative.


\subsection{Overview}
\label{Overview}

We have shown above that the neutral part of the absorbing cloud covers the source
of the intrinsic quasar continuum completely but only partly the broad emission line region.
We therefore are interested in which fraction of the BLR is covered by the cloud.

For this, we estimate the fraction of the total flux emitted by the continuum source
at the location of the emission lines. This is done using the spectral shape of the intrinsic
continuum as shown in Fig. 1.
It is obtained by extrapolating the
continuum below the emission lines by splines functions fitted in the regions devoid of any emission or
absorption lines.
We thus can estimate both the fluxes from the BLR, F$_{\rm BLR}$, and the continuum. Using the values
of LFR derived in the previous Sections and the fact that the cloud covers the continuum completely,
it is straightforward to derive the covering factor of the BLR for different emission lines $\rm f = 1 - (LFR\cdot F_{total})/F_{BLR}$.
Results are summarized in Table~\ref{table_cf}.

When estimating the errors, we have taken into account the errors on LFR but also
a $\sim$20-30\% uncertainty on the QSO intrinsic continuum placement below the emission lines.
Note that for further discussion we use only covering factors of C\,{\sc iv} and Lyman-$\alpha$ emission
lines.

\begin{table*}
   \begin{tabular}{cccccc}
   \hline
   \hline
      &  O~{\sc vi} & Lyman-$\alpha$ & N$\,${\sc v} & C$\,$IV \\ 
   \hline
   H$_2$   &  $0.80^{+0.07}_{-0.11}$  &   &   &   \\ 
   C\,{\sc i}   &   & $0.38^{+0.16}_{-0.32}$ & $0.56^{+0.14}_{-0.21}$ & $0.47^{+0.07}_{-0.12}$ \\ 
   O\,{\sc i}   &   & $0.93^{+0.02}_{-0.02}$ &   &   \\ 
   Si\,{\sc ii} &   & $0.91^{+0.02}_{-0.03}$ &   &   \\ 
  \hline
   \end{tabular}
   \caption{Covering factors of different emission line regions (columns) by different species (rows)
after taking into account the fact that the QSO intrinsic continuum is fully covered.
   We note that the most reliable measurement is for C\,{\sc iv} emission line (see text).}
   \label{table_cf}
\end{table*}

\section{Physical conditions in the DLA system}

If we could estimate the size of the intervening {H$_2$-bearing} molecular cloud, we could estimate
the size and, possibly, the structure of the QSO BLR using the covering factors
derived in the previous Section. Therefore, we will first
study the physical conditions in the gas.


\subsection{Ionization structure of the $z=2.3377$ absorption cloud}
\label{Ioniz_struc}
\begin{figure*}
            \includegraphics[width=0.95\textwidth,clip]{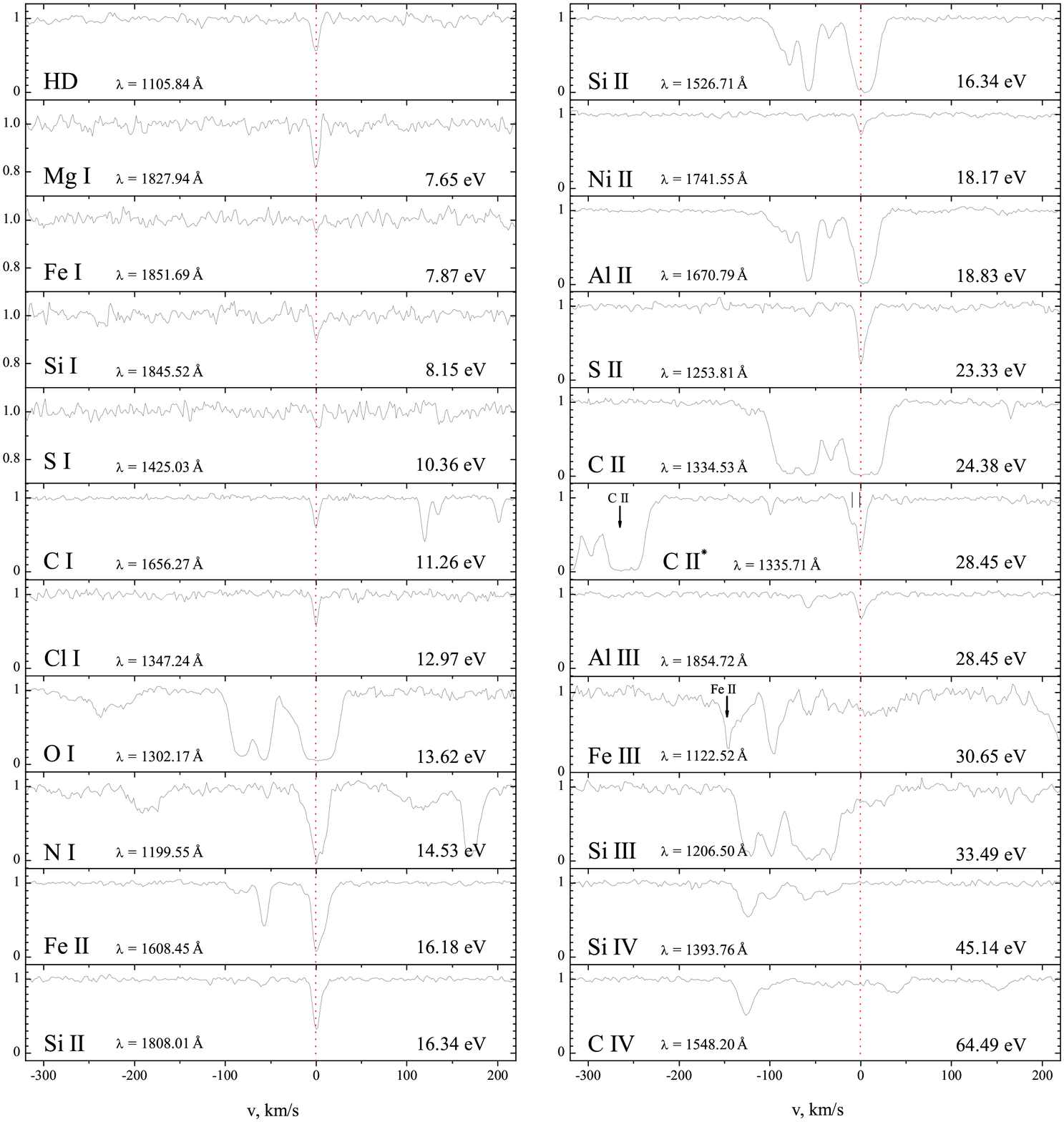}
            \caption{Absorption profiles of different species in the $z=2.3377$ DLA system
towards \qso. The origin of the velocity scale is taken at the redshift
of the H$_2$ component, $z=2.33771$.}
            \label{Ioniz}
\end{figure*}

The absorption profiles of different species are shown in Fig.~\ref{Ioniz} ranked following the
ionization potential of the corresponding ion.
The {H$_2$-rich} component is traced by HD and neutral species Mg~{\sc i}, Fe~{\sc i}, Si~{\sc i},
S~{\sc i}, C~{\sc i} and  Cl~{\sc i}. The latter {species} is closely tied up to H$_2$ by charge
exchange reaction processes.
We tentatively detect Fe\,{\sc i} and Si\,{\sc i} absorptions, associated with the molecular component.
The measured column densities $\log(N)$ are 11.86$\pm$0.09 and 12.68$\pm$0.18 for Si\,{\sc i} and Fe\,{\sc i},
respectively. These ions are detected only in a few QSO absorption systems
(D'Odorico 2007; Quast, Reimers \& Baade 2008) and indicate the presence of cold and well shielded gas.
Neutral sulphur is rarely seen in QSO absorption systems but it is seen in {five} absorption systems with
associated CO detection (Srianand et al. 2008; Noterdaeme et al. 2009; Noterdaeme et al. 2010; Noterdaeme et al. 2011). We therefore carefully searched for CO absorptions in the \qso\,\, spectrum. Unfortunately, we could place only
a 3\,$\sigma$ upper limit on CO {column density from the non-detection in the three strongest band at
1447~\AA, 1477~\AA\, and 1509~\AA: log~$N$(CO)~$<$~12.6.} The singly and twice ionized species span about
150~km~s$^{-1}$ {bluewards} of the {H$_2$-bearing} molecular component.

{The ionization structure indicates that a {large fraction of the metals could be} associated with the
molecular component. This is the case of low ionization species (E$<13.6$\,eV) but also of higher ionization
species (see for example S\,{\sc ii} and Si\,{\sc ii} absorption profiles). This indicates that the cloud
is concentrated around the central H$_2$-bearing component.}


\subsection{Metal content}

We have fitted all the absorption lines which are not strongly saturated using Voigt-profiles, derived
total column densities and calculated metallicities
as [X/H]~=~log~$N$(X)/$N$(H~{\sc i})~$-$~log~[$N$(X)/$N$(H~{\sc i})]$_{\odot}$ with
solar metallicities taken from Lodders (2003).
Results are given in Table~\ref{table_metal} and are summarised in Fig.~\ref{Metal}.
The mean metallicity as indicated by the Sulfur metallicity which has smallest error
is [S/H]~=~$-$1.32$\pm$0.12.
{It is remarkable that, within errors, [S/H]~$\sim$~[Si/H], which indicates little depletion of Si.

Note that [Cl\,{\sc i}/H]~$\sim$~[S/H] seems at odd.
As Cl~{\sc i} is coupled with H$_2$ via charge exchange reactions (Jura 1974), Cl~{\sc i} comes from the
H$_2$-bearing component. Since we measure a high Cl~{\sc i} column density, this H$_2$
bearing component must have a large molecular fraction ($>$0.25, see Abgrall et al. 1992).

Therefore, in principle the Cl metallicity could be much higher since we should divide
the Cl~{\sc i} column density by 2$\times$$N$(H$_2$) and not by the total hydrogen column density
($N$(H\,{\sc i})+2$N$(H$_2$)), as we have done in Table~\ref{table_metal}. This would give a metallicity close
to solar in the H$_2$ bearing component [Cl\,{\sc i}/2H$_2$]$<$~$-$0.27$\pm$0.17. This is an upper limit
as the cloud could be partially molecular only.

It is likely that most of the Si~{\sc ii} is to be found associated with the molecular
component (see Fig.~\ref{SiandOfig}, $\lambda$=1808\,\AA\,line). This further suggests that the cloud is concentrated
around the clumpy central component of higher metallicity.

Relative abundance of nitrogen to $\alpha$ elements, i.e. [N/S], [N/Si] is $\sim -1$, that is consistent
with typical low nitrogen metallicity in DLAs (Petitjean, Ledoux \& Srianand 2008; Pettini et al. 2008).
Iron, Nickel and Manganese are observed {to be} depleted by a factor $\sim$5 relative to Sulfur.
This moderate depletion probably onto dust-grains supports the idea {that the presence of} even little
amount of dust favors the formation of molecular hydrogen.


\begin{table}
    \centering
    \begin{tabular}{|c|c|c|}
        \hline\hline
        Species & log $N$ & [X/H]$^{\rm a}$ \\
        \hline\hline
        N~{\sc i}	& $14.54\pm 0.22$ & $-2.23\pm 0.24$ \\
        Mg~{\sc ii}	& $15.33\pm 0.24$ & $-1.16\pm 0.25$ \\
        Si~{\sc ii}	& $15.06\pm 0.05$ & $-1.42\pm 0.09$ \\
        P~{\sc ii}	& $12.86\pm 0.24$ & $-1.54\pm 0.25$ \\
        S~{\sc ii}	& $14.81\pm 0.09$ & $-1.32\pm 0.12$ \\
        Cl~{\sc i}	& $12.97\pm 0.14$ & $-1.23\pm 0.16$ \\
        Ar~{\sc i}	& $13.86\pm 0.22$ & $-1.63\pm 0.24$ \\
        Mn~{\sc ii}	& $12.22\pm 0.08$ & $-2.22\pm 0.11$ \\
        Fe~{\sc ii}	& $14.44\pm 0.08$ & $-1.97\pm 0.11$ \\
        Ni~{\sc ii}	& $12.81\pm 0.04$ & $-2.35\pm 0.09$ \\
                \hline
            \end{tabular}
            \caption{Column densities (in units of cm$^{-2}$) and metallicities relative to solar in the $z=2.33771$ DLA system.
            $\rm ^a$\,\,Relative to solar. Solar abundances are taken from Lodders (2003).}
            \label{table_metal}
\end{table}


\begin{figure}
            \includegraphics[width=0.45\textwidth,clip]{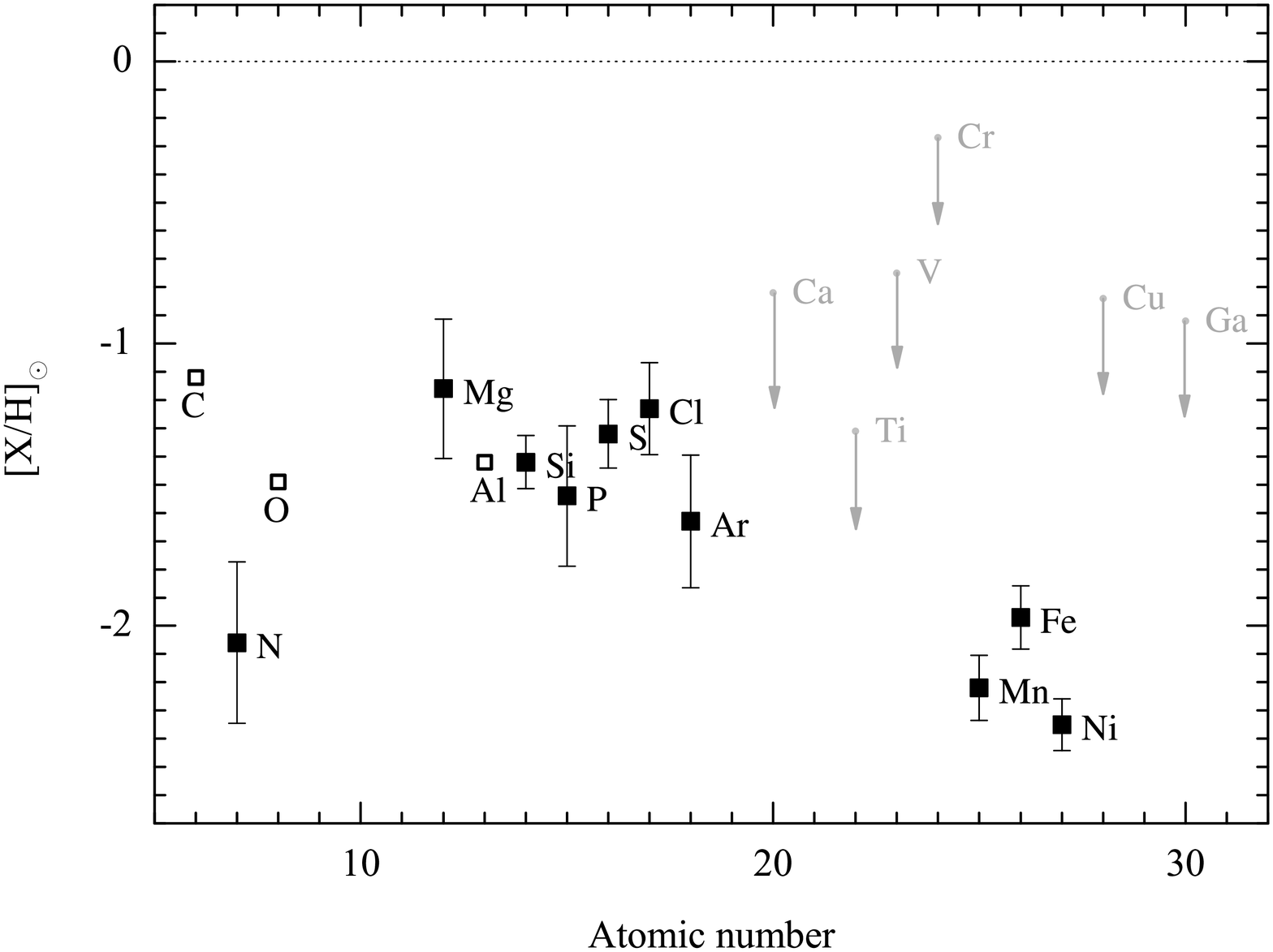}
            \caption{Metallicities relative to solar in the $z=2.33771$ absorption system toward \qso.
Empty squares show C, O and Al metallicities derived from saturated lines and should {be} considered as lower limits.}
            \label{Metal}
\end{figure}


\subsection{Number densities}
\label{Phys_cond}


\subsubsection{C\,{\sc i} Fine structure}
\label{fineCI}

In Section~\ref{CIpc} we have measured the column densities of C\,{\sc i} {atoms} in the fine structure
levels of the ground state. The balance between the different level populations can be used to estimate the
number density in the gas.
{Radiative pumping of C\,{\sc i} excited states is probably not important since to explain the
measured column density ratios the UV radiation field should be about 50 times higher than the mean Galactic value
(see the discussion in Silva \& Viegas (2002) and Noterdaeme et al. (2007)). Such a high UV radiation
field is very unlikely for this one component system in which low ionization species dominate. We also assume the
CMBR temperature to be $T_{\rm CMBR}$~=2.73~$\times(1+2.3377)$~K, following standard
cosmology model. Under these two assumptions, and considering a homogeneous cloud, the relative populations
of the C\,{\sc i} levels depend only on the number density and the temperature. Collision coefficients
were taken from Schr\"oder et al. (1991), assuming that the main collisional partner is H$_2$.
We take for the temperature in the cloud the value, $T_{\rm kin}$~=~67$\pm$11~K, derived from the analysis
of the molecular hydrogen ortho-para ratio (Ivanchik et al. 2010). Confidence contours for
$n_{\rm H_2}$ are shown on Fig.~\ref{CIfine}. The best value for the density is $n_{\rm H_2}$~=~105$^{+27}_{-24}$~cm$^{-3}$
with 1$\sigma$ uncertainty.


\begin{figure}
            \includegraphics[width=0.45\textwidth,clip]{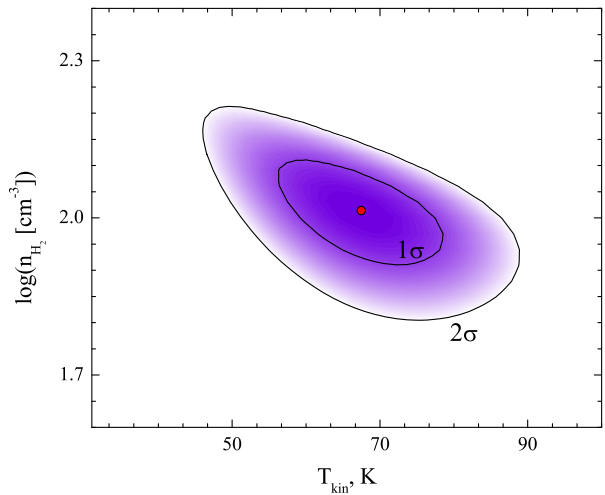}
            \caption{Confidence contours for the physical conditions derived from the relative populations
of the C\,{\sc i} fine structure levels and the populations of the {first two} rotational levels
of H$_2$.
}
            \label{CIfine}
\end{figure}


\subsubsection{HD rotational levels}
\label{HDlevels}

We can derive an upper limit on the J~=~1 HD column density from the most prominent absorption lines
shown in Fig.~\ref{HDJ1}. All other HD J~=~1 transitions are partly or fully blended.
We find log~$N$(HD, J=1)~$<$~14.1. The measured {$N$(HD, J=1)/$N$(HD, J=0)} upper limit is in agreement with
what is expected in {HD/H$_2$} molecular clouds (Le Petit, Roueff \& Le Bourlot 2002) and can be used to derive an upper limit on the number
density in the cloud. For this, we follow the analysis of Balashev, Ivanchik \& Varshalovich (2010) who have detected recently HD J~=~1
absorptions at $z$~=2.626 towards J\,0812+032. Likewise the C\,{\sc i} analysis, we can neglect radiative pumping. In spite of
excitation energies of HD levels being higher than those of C\,{\sc i} (which makes collisions less effective), self-shielding
processes become rapidly important for HD. We assume the temperature in the cloud is given by the H$_2$ excitation temperature,
$T_{\rm kin}$~=~67$\pm$11~K and that H$_2$ is the main collisional partner (with collisional coefficients taken from
Flower et al. (2000)). The upper limit on the density is $n_{\rm H_2}<$160~cm$^{-3}$, in agreement with the value derived from C\,{\sc i}.


\begin{figure}
            \includegraphics[width=0.45\textwidth,clip]{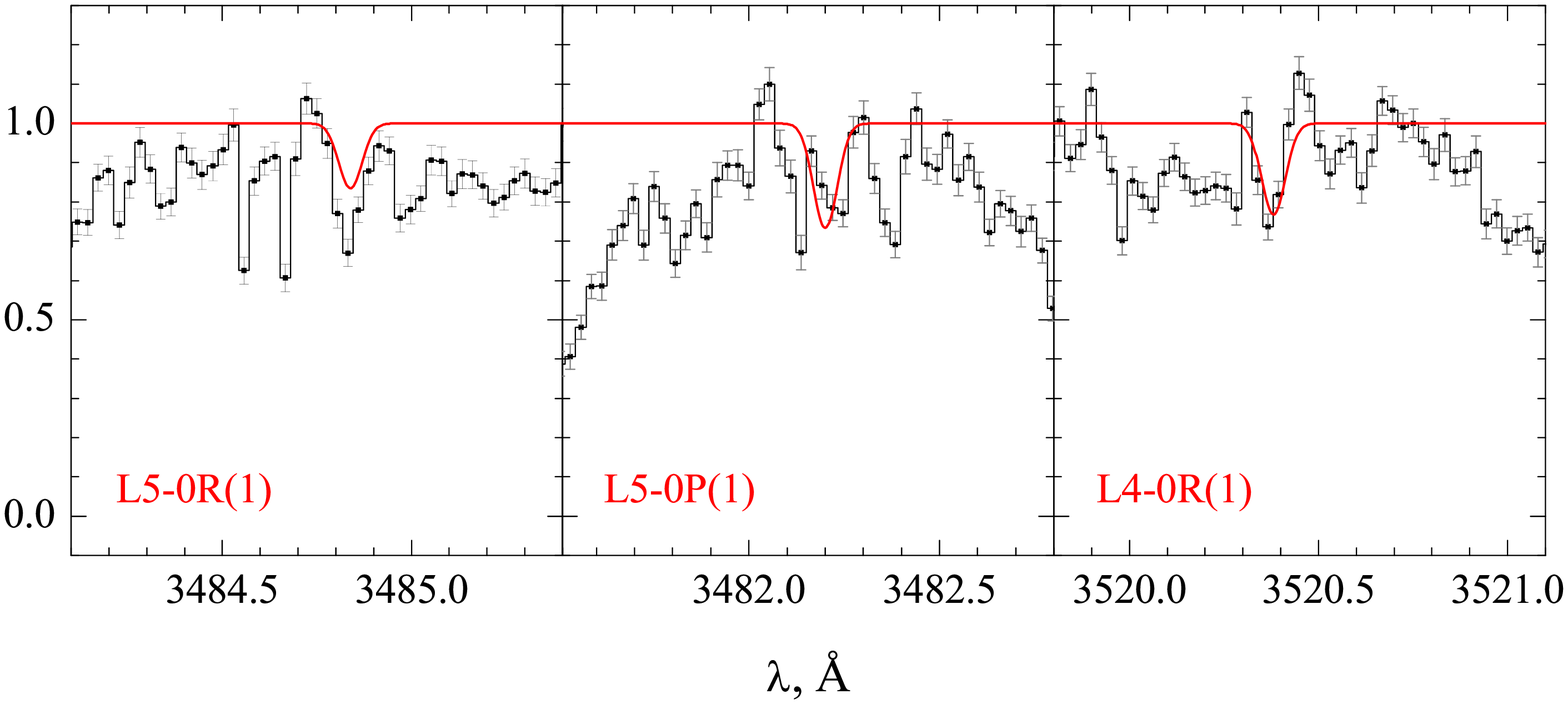}
            \caption{{Expected location of} HD(J=1) {lines from the} $z=2.33771$ absorption system toward \qso.}
            \label{HDJ1}
\end{figure}


\subsubsection{C\,{\sc ii} fine structure}
\label{fineCII}

Singly ionized carbon is mostly located in a region surrounding the {H$_2$-rich} molecular core of the cloud.
Therefore, from the excitation of its fine structure level it is possible to estimate the number density
in the envelope of the molecular{-rich} cloud.
From the $\lambda=1335.7$ {\AA}\, line (see Fig.~\ref{Ioniz}), we measure
log~$N$(C\,{\sc ii}$^*$)~=~13.92$\pm$0.15.
Unfortunately it is not possible to measure $N$(C\,{\sc ii}) directly as the C\,{\sc ii}$\lambda$1334 absorption
line is strongly saturated (see Fig.~\ref{SiandOfig}) and we have to estimate it indirectly (see Srianand, Petitjean \& Ledoux 2000).
It is known, that the metallicity of the $\alpha$-chain elements is enhanced compared to carbon by a factor
of about two, when the mean metallicity is low (i.e. $<-1.0$). Therefore, we consider that $Z$(C)~=~$Z$(Si)/2 {and}
derive {log~$N$(C\,{\sc ii})~=~15.6$\pm$0.3.}
In a neutral medium, the upper level of C\,{\sc ii} is predominantly excited by collisions with hydrogen (Silva \& Viegas 2002;
{Srianand et al. 2005}).
This assumption is supported by the velocity structure of O\,{\sc i} which traces the neutral gas and is present over
the whole C~{\sc ii} absorption profile.
The probability {of the ground state fine structure transition 3/2 $\to$ 1/2} is $A_{21}\sim 2.29 \times 10^{-6}$~s$^{-1}$
and the excitation rate with atomic hydrogen is taken from Silva \& Viegas (2002).
For the range in temperature $T$~$\sim$~100~$-$~2000\,K, we estimate the density of the neutral region,
$n_{\rm HI}$~=~32$^{+34}_{-21}$~cm$^{-3}$.
This is smaller but not very different from the density found for the molecular{-rich} core.

\section{Discussion}
\label{discussion}
\noindent

In this Section we discuss several interpretations to explain partial coverage of
QSOs by an intervening cloud:
binary quasars, gravitational lensing and transverse dimensions. {These explanations are schematically pictured in
Fig.~\ref{assumptions}}.
In our case and as shown in previous Sections,
the continuum source is fully covered by the molecular cloud when the BLR seems to be partially
covered by the dense and neutral part of the cloud but also by part of the singly ionized species.

\begin{figure}
		\centering
            \includegraphics[width=0.40\textwidth,clip]{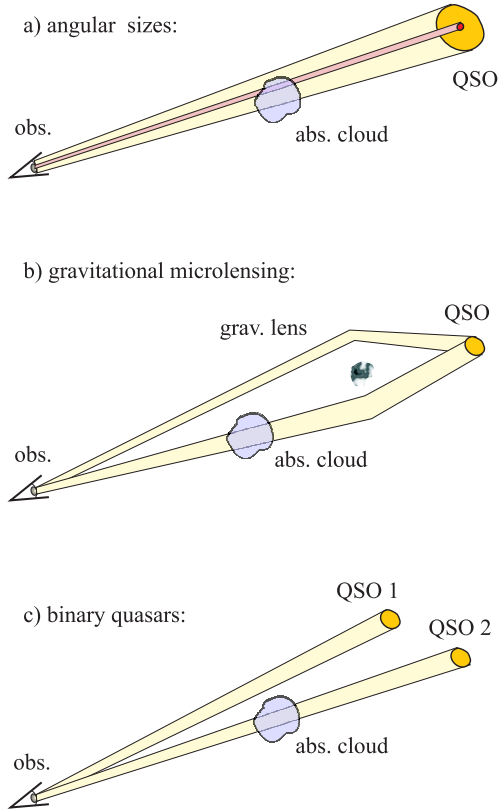}
            \caption{Illustration of the possible explanations for the partial coverage of QSOs by an
intervening absorption system.}
            \label{assumptions}
\end{figure}


\subsection{Binary quasars}
\label{DoubleQSO}
\noindent

Binary quasars are thought to be the consequence of a galaxy merging event and are observed with separations as small as approximately 10~kpc (Hennawi et al. 2006; Foreman, Volonteri \& Dotti 2009; Rodriguez et al. 2006; Vivek et al. 2009).
At redshift $z\sim 2$ binary quasars with transverse distance $\lesssim 3\, {\rm kpc}$ would be unresolved by UVES observations.
Partial coverage can be seen in case the background source is a binary quasar if the absorbing cloud covers one of the quasars and not the other.
However, in such a situation, the intrinsic continuum source would not be covered completely.
This is not the case of the z=2.3377 absorption system towards Q~1232+082 which covers the intrinsic continuum completely.


\subsection{Gravitational Microlensing}
\label{Microlensing}
\noindent

Gravitational lensing (especially microlensing) is one of the explanations for partial coverage. In the simplest
case of the quasar being macro-lensed by a galaxy or a galaxy cluster, there is no difference in the QSO
image spectra. We therefore expect partial coverage of both the BLR and the intrinsic continuum. Since in the present case,
the cloud is very small, the contrary would need a very special configuration.

Observations {by Sluse et al. (2007) and Wucknitz et al. (2003)} indicate that additional microlensing by stars or objects
with mass $\sim 10^4 M_{\odot}$ can enhance part of the spectrum (continuum and/or a fraction of the BLR)
in only one image. If the continuum is strongly enhanced compared to the BLR in the image that is not covered
by the cloud, then partial coverage in the continuum could stay unnoticed.
However, this explanation would request again a very special situation.

Indeed, if we consider the spectrum as a superposition of a continuum and emission lines,
$F = F_{\rm em} + F_{\rm c}$, then in case two images are macrolensed
with factors $M_1$ and $M_2$ and only one image is microlensed with a factor $\mu$,
the fluxes in the two images are:

\begin{equation}
        F_1  = M_1 \cdot F_{\rm em} + M_1 \cdot F_{\rm c}
\end{equation}
\begin{equation}
        F_2  = M_2 \cdot F_{\rm em} + M_2 \cdot \mu \cdot F_{\rm c}.
\end{equation}

In the simple case of the absorption system being located in front of
the microlensed image and totally absorbing it, the measured covering
factors in the continuum, ${f}_{\rm c}$, and in the emission line, ${f}_{\rm em}$, will be:

\begin{equation}
       {f}_{\rm c} = \left(\frac{F_2}{F_1+F_2}\right)_{\rm c}  = \frac{M_2\cdot \mu}{M_2\cdot\mu+M_1}
\end{equation}
\begin{equation}
      {f}_{\rm em} = \left(\frac{F_2}{F_1+F_2}\right)_{\rm em}  = \frac{M_2}{M_1+M_2},
\end{equation}

In the case of the absorber in front of Q~1232+082, we observe
$f_{\rm em} \sim 0.5$ and  $f_{\rm c} \gtrsim 0.98$ (from zero level correction see Sect.~\ref{correction}). This would imply $M_1 \sim M_2$
but more importantly requires a probably unreasonably large microlensing effect $\mu \gtrsim 50$.


\subsection{BLR kinematics and size in \qso}
\label{BLRsize}

We discuss here the information we can derive on the size of the BLR from the
observation and analysis presented above. For this we use over-simplified models
considering spherical geometry. This is speculative but given
the lack of information on the exact geometrical and kinematical structure of the
BLR (Denney et al. 2010), may still be interesting.

The most probable explanation for partial coverage of the \qso\,\, BLR by the $z=2.3377$ absorption system is that
the transverse size of the cloud is smaller than the projected size of the BLR.
Following the standard paradigm for Active Galactic Nucleus (AGN), the QSO emission takes place in
different regions of different sizes. Therefore, if the transverse extent of the intervening cloud
is smaller than the maximum extent of the background source, partial coverage can happen.
The inner part of the accretion disk produces the continuum emission.
Linear size  of the disk is of the order of $l_{\rm disk} < 10^{15}\,$cm (Dai et al. 2010 and references within), corresponding to
$\theta_{\rm disk} \approx 10^{-6}$~mas at $z=2.57$ (with $h_{100}=0.72$, $\Omega_M=0.27$, $\Omega_{\Lambda} = 0.73$).
The linear size of the BLR is several orders of magnitude larger than the size of the accretion
disk and of the order of $\sim 1\,$pc (e.g. Wu et al. 2004; Kaspi et al. 2005).
Thus the angular size of the BLR is limited to about $\theta_{\rm BLR} \lesssim 10^{-2}$~mas.
In addition, the BLR is supposed to be stratified, the low ionization lines having larger extent
than the high-ionization ones (Peterson 1993).

We have estimated the number densities in the cloud:
$n_{\rm H2}$~$\sim$~110~cm$^{-3}$ for the {H$_2$-bearing} molecular core and $n_{\rm H}$~$\sim$~30~cm$^{-3}$ for the neutral envelope.
Given the column density, $N$(H$_2$)~=~$4.78\pm0.96$~$\times$10$^{19}$~cm$^{-2}$,
this gives $l_{\rm H2}$~$\sim$~$0.15^{+0.05}_{-0.05}$~pc for the linear size of the {H$_2$-bearing} molecular
core along the line of sight. This should also correspond to the size of the C~{\sc i}
cloud. Indeed the kinematical structure of the H$_2$, HD, C~{\sc i} and Cl~{\sc i} absorption features show a single
narrow component at the same position (see Fig.~10). Remember also that Cl~{\sc i} and H$_2$ are tied up
(see e.g. Noterdaeme et al. 2010).
With the assumption that the transverse size of the cloud is of the same order of magnitude
as the longitudinal size, we find that this is small enough to explain the partial coverage of
the BLR. To estimate the BLR size we assumed that the cloud and BLR are spherical in shape,
and used
\begin{equation}
  L_{\rm BLR} = L_{\rm abs} \times K_{\rm cosm} \times K_{\rm geom},
\end{equation}
where $L_{\rm BLR}$ and $L_{\rm abs}$ are the transverse sizes of the BLR and the absorber, respectively.
$K_{\rm cosm}$ is a cosmological correction factor due to the fact that the QSO and the absorption cloud
have different redshifts. In the standard cosmological model and because the redshifts are similar
($z_{\rm em}$~=~2.57 and $z_{\rm abs}$~=~2.3377),
this factor is about 1. $K_{\rm geom}$ is a coefficient that depends on the alignment of the two objects
and therefore on the distance, $l$, between the two projected centers.
Given the dimension of the absorber, we can calculate the most probable dimension of the
background source for an observed covering factor. Since the absorber and the background sources
are not related, the alignment is random and the probability that $l=l_0$ is proportional
to $l_0$.
Therefore, one can calculate the probability distribution function for $K_{\rm geom}$
and obtain the best value of $K_{\rm geom}$ and its errors.
Using the measured covering factor, $\sim$0.47, for the C\,{\sc iv} emission line, we derived
this way a size of the C\,{\sc iv} BLR $R_{\mbox{C\,{\sc iv}}}$~$\sim$~$0.18^{+0.08}_{-0.11}$~pc.
A schematic representation of this situation is illustrated on Fig.~\ref{sizes}.

We can use the size$-$luminosity relationship derived from reverberation mapping  (Peterson et al. 2005,
Kaspi et al. 2007) to estimate the size of the C\,{\sc iv} BLR to be compared with our results.
We estimate the luminosity of Q~1232+082, $\lambda$$L_{\lambda}$(1350~\AA) (erg/s), to be
$\sim$10$^{46.9}$. Using Fig.~6 of Kaspi et al. (2007) we derive a C~{\sc iv} lag of the order
of 300~days, corresponding to about 0.26~pc.
Note that recently, the study of differential microlensing in
Q~2237+0305 (Sluse et al. 2010) yielded similar results for the size of the C\,{\sc iv} BLR, $0.06^{+0.09}_{-0.04}$~pc.
The same typical size is
obtained from the study of GeV break in blazars (Poutanen \& Stern 2010).

\begin{figure}
            \includegraphics[width=0.45\textwidth,clip]{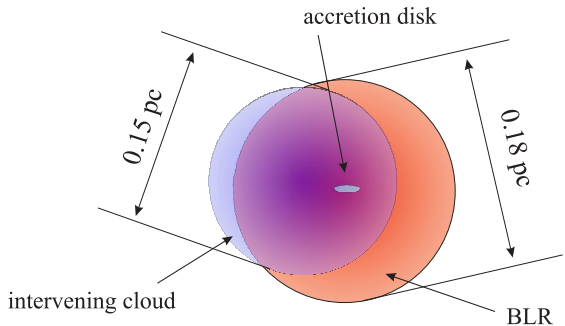}
            \caption{Partial coverage of \qso\,\, BLR by an intervening absorption cloud at $z=2.3377$.
Geometry is assumed spherical.
The estimated size of the BLR corresponds to the C~{\sc iv} emission line region covered at 48\%.}
            \label{sizes}
\end{figure}

On the other hand, {for the neutral phase, we have} $N$(H~{\sc i})~=~7.94$\times$10$^{20}$~cm$^{-2}$
and therefore $l_{\rm HI}$~$\sim$~$8.2^{+6.5}_{-4.1}$~pc. Remember that the neutral phase of the cloud
as traced by O~{\sc i} and Si~{\sc ii} covers only $\sim$94\% of the Lyman-$\alpha$ emission.
The large size we derive for the neutral cloud extent seems in contradiction with
the above small size of the C~{\sc iv} BLR. However, it must be reminded that the
O~{\sc i} and Si~{\sc ii} are located nearly exactly on top of the Lyman-$\alpha$ emission line.
This would mean that the Lyman-$\alpha$ BLR is probably fully covered but that the Lyman-$\alpha$
emission extends well beyond the extent of the C IV emitting BLR. In turn this means that the extended
Lyman-$\alpha$ emission corresponds to about 6~\% of the flux at the peak of the
emission. Note that the velocity difference between O~{\sc i}\, $\lambda$1302 and
Si~{\sc ii}\, $\lambda$1304 is $\sim$500~km/s. Therefore, as can be seen
on Fig.~8, right panel, the kinematics of the region which is not covered must be of this order.
Unfortunately, we cannot go farther in our derivation of the extent of the Lyman-$\alpha$ emission.

The determination of the BLR size $R_{\rm BLR}$ allows us to estimate the mass of the central black-hole, associated with \qso\,\,
assuming that the BLR is virialized:
\begin{equation}
    M_{\rm BH} = \phi\frac{R_{\rm BLR}\Delta V^2}{G},
\end{equation}
where $\phi\approx 5.5$ is a scale factor (Peterson et al. 2004) and $G$ is the gravitational constant.
We estimate the width of the C\,IV emission line $\Delta V = 2463^{+12}_{-24}$\,{km/s}, which is less than the
value $\sigma_v \approx 2780$\,{km/s} derived by Vanden Berk et al. (2001).
We find $M_{\rm BH}$~=~6.8$^{+4.1}_{-4.5}$$\times$10$^{8}$~M$_{\odot}$.
This is slightly lower than the mean BH mass derived by Vestergaard et al. (2008) for quasars of this luminosity
at this redshift but well within the overall scatter.


\section{Conclusion}
\noindent
\label{conclusion}

The analysis of H$_2$, C\,{\sc i}, O~{\sc i} and Si~{\sc ii} absorption lines
from the molecular DLA system at $z_{\rm abs}$~=~2.3377 toward \qso\,
has shown that the intervening absorbing cloud is not covering
the background source totally.
We used a curve-of-growth and a profile fitting analysis to
estimate the partial covering factor.
The different methods yield covering factors of the {H$_2$-bearing} core
(as traced by H$_2$ and C~{\sc i}) to be
$\approx$~48, 66 and 75~\% for the C~{\sc iv}, C~{\sc iii} and Lyman-$\beta$-O~{\sc vi}
emission lines {whilst} the QSO intrinsic continuum is covered completely.
The O~{\sc i}\,$\lambda$1302 and Si~{\sc ii}\,$\lambda$1304 absorptions
cover only $\sim$94\% of the Lyman-$\alpha$ emission.

According to the generally accepted model of AGNs, broad emission lines are
emitted by warm and highly ionized gas located in the BLR with transverse dimension
of the order of $\sim$1~pc. The quasar continuum is produced by the inner part of the
accretion disk, which linear size is several orders of magnitude less than the BLR size.
Thus the most probable explanation of the observed partial coverage
is the comparable angular sizes of the BLR and the compact {H$_2$-bearing} absorption cloud.
The fact that the continuum is completely absorbed makes
other explanations such as the presence of a binary quasar or
gravitational lensing less plausible.

We derived the linear extent of the H$_2$-bearing cloud and neutral envelope to be $l_{\rm H_2}$~$\sim$~$0.15^{+0.05}_{-0.05}$~pc
and $l_{\rm HI}$~$\sim$~$8.2^{+6.5}_{-4.1}$~pc, respectively.
Assuming that the {H$_2$-bearing} component is spherical in shape, we estimate the size of the C~{\sc iv}
BLR to be $\sim$0.2~pc. The large size we derive for the neutral cloud extent together with the covering
factor of $\sim$94\% of the Lyman-$\alpha$ emission means that the Lyman-$\alpha$
BLR is probably fully covered but that the Lyman-$\alpha$ emission extends well beyond the extent of the C IV emitting BLR.
In turn this means that the extended Lyman-$\alpha$ emission corresponds to about 6~\% of the
flux at the peak of the emission (over $\sim$500~km/s).
Assuming the C~{\sc iv} BLR is virialized, we derive the mass of the central
$M_{\rm BH}$~=~6.8$^{+4.1}_{-4.5}$$\times$10$^{8}$~M$_{\odot}$

Partial coverage of the background source by intervening clouds has been observed an used to derive the
radius of the clouds when the background source is the combination of several gravitational images
(Petitjean et al., 2000a; Ellison et al., 2004).
This is the first time that partial coverage of a QSO BLR by an intervening cloud is reported.

This kind of situation tests the kinematics of the BLR. Indeed we see that the covering factor
depends of the position of the absorption on top of the emission line.
This can be due to the presence of inflows or outflows in BLR (Denney et al. 2009) which could be covered
differently or
the BLR could have a disk-like structure
implying the projected velocity is correlated with the spatial position.
Obviously it will be difficult to {disentangle} the different kinematics. However, it may be interesting to
dramatically increase the quality of spectra in which an intervening compact absorption system is observed.
This may reveal that partial covering factor is not unusual and could be used to constrain the BLR structure.

\vspace{2mm}{\footnotesize {\rm Acknowledgments.} This work was
supported in part by a bilateral program of the Direction des Relations
Internationales of CNRS in France, by the Russian Foundation for
Basic Research (grant 11-02-01018a), and by a State Program
``Leading Scientific Schools of Russian Federation'' (grant
NSh-3769.2010.2), by Ministry of Education and Science of Russian Federation
(contract \# 11.G34.31.0001 with SPbSPU and leading scientist G.G. Pavlov). PPJ and RS acknowledge support from the Indo-French Centre for the Promotion of Advanced Research under the programme No.4304--2.
SB thanks Dynasty Foundation and A.M. Krassilchtchikov for help with cluster computations.}

\label{lastpage}

\bsp

\end{document}